
\magnification=1200
\hsize 15.0 true cm
\vsize 23.0 true cm
\def\hrf{\hrulefill}
\def\A{{\rm A}}
\def\c{\centerline}
\def\v{\vskip 1pc}
\def\ej{\vfill\eject}
\def\r{\vec r}
\def\ra{\rangle}
\def\la{\langle}

\def\tr{{\rm tr}}
\def\half{{1 \over 2}}
\def\da{\dagger}
\def\pup{{\rm p} \uparrow}
\overfullrule 0 pc
\
\vskip 1.5pc
\parindent 3pc\parskip 1pc
\c{\bf Spin content from skyrmions with parameters
fit to baryon properties}
\vskip 4 pc
\c{G. K\"albermann$^{\rm a}$ and J.M. Eisenberg$^{\rm b}$}
\v
\c{$^{\rm a}$ {\it Rothberg School for Overseas Students and Racah Institute
of Physics}}
\c{\it Hebrew University, 91904 Jerusalem, Israel}
\vskip 0.5 pc
\c{$^{\rm b}$ {\it School of Physics and Astronomy}}
\c{\it Raymond and Beverly Sackler Faculty of Exact Sciences}
\c{\it Tel Aviv University, 69978 Tel Aviv, Israel}

\vskip 4pc
\noindent {\bf Abstract:-}  Earlier work reported on the existence of a
term within a generalized skyrmion approach that yields appreciable
spin content for the proton.  Unfortunately there is no accessible
experiment that can fix the coefficient of this term directly; plausible
but highly uncertain values for it gave a result for the spin content
loosely consistent with the currently measured
$\Delta\Sigma = 0.27\pm 0.13.$  We here attempt
to narrow the range of values for this coefficient by performing
global fits to all the parameters of the generalized Skyrme lagrangian
while
requiring reasonable results for the baryon octet and decuplet masses and
octet magnetic moments.  This requirement fixes the coefficient
loosely, and we find that parameter sets that fit the baryon
masses and magnetic moments yield proton spin content near
$\Delta\Sigma \sim 0.15.$

\vfill

\noindent January, 1995.

\ej

\baselineskip 15 pt
\parskip 1 pc \parindent 3pc
\c{\bf 1. Introduction}

When the original measurement [1] of the spin content of the proton
yielded a result consistent with zero it was deemed to be a
merit of the Skyrme model [2] that it also suggested a vanishing value
for this quantity.  Already in the early work [3] that showed this
it was pointed out that some ways of breaking flavor SU(3) could provide
nonzero spin content $\Delta\Sigma,$ but at least in the approach chosen
there $\Delta\Sigma$ proved to be negative.  Thus when more recent
measurements [4] gave, for all proton data,
$\Delta\Sigma = 0.27 \pm 0.13$ it became urgent
to find sources of positive spin content in the general skyrmion
approach if that model were to survive [5-9].

Recently a term in a generalized Skyrme lagrangian has been
found [10] that can yield sizable positive proton spin content thus
making it straightforward to work with a skyrmion that includes
this feature.  The term ${\cal L}_{6,1}$ in question [see eq. (1)]
contains six derivatives of
the skyrmion field.  For flavor SU(2) it reduces to a form previously
put forth [11,12] as a possible device for stabilizing the skyrmion
in place of, or alongside, the usual four-derivative term originally
proposed by Skyrme [2].  Thus the presence of this term automatically
leaves unchanged the
many successful features of the model in SU(2).  On the other hand
for flavor SU(3) it yields new results including the improved proton
spin content.  Unfortunately a direct determination of the coefficient
$\epsilon_1$ of this term cannot be made from particle decay rates, the
relevant transitions either vanishing identically in this model
(viz., $\eta' \rightarrow 5\pi$), being kinematically forbidden
(viz., $\eta' \rightarrow \eta + 4 \pi$), or being hopelessly
difficult to measure (e.g.,  $\pi\pi \rightarrow \pi\pi \eta \eta'$).
Thus ref. [10] contented itself with suggesting plausible values for
this new coefficient along with the older and better known coefficients
for the other terms in the generalized Skyrme lagrangian.  These then
led to $\Delta\Sigma \sim 0.3$ to 0.4 which is roughly consistent with
current experiment [4].

It is our purpose here to try to narrow the range of acceptable values
for $\epsilon_1$ by carrying out an extensive fit to the masses of the
baryon octet and decuplet.  The centroids and splittings of the masses
for the nonzero strangeness members of these multiplets are sensitive
to $\epsilon_1$ since ${\cal L}_{6,1}$ influences the SU(3) skyrmion
mass and moments of inertia [10].  We shall also use the magnetic
moments of the octet as a further check on the resulting values for
$\epsilon_1$ as these moments are usually reliably calculated in the
skyrmion approach.  We do not use here the values for charge
and magnetic radii which are known only for the nucleon.
\v

\c{\bf 2. Brief review of the formalism}

Since the relevant formalism of Yabu and Ando [13] for the flavor
SU(3) skyrmion is well known and the modifications required for the
inclusion of the
six-derivative term with appreciable spin content have already
been presented in ref. [10], we sketch only the essential
features of the development required to fix notation and set the
background of the problem.  The generalized skyrmion is
described by the lagrangian\footnote{$^1$}{We note that a general
analysis of effective chiral lagrangians with six-derivative terms
has recently become available [14].}
$$\eqalign{{\cal L} & = {\cal L}_2 + {\cal L}_{4}
+ {\cal L}_{6,1} + {\cal L}_{6,2} + {\cal L}_{SB}\cr
& = -{F_\pi^2 \over 16} \tr(L_\mu  L^\mu )
+ {1 \over 32 e^2} \tr[L_\mu,L_\nu]^2 \cr
& - \epsilon_1 {g_\omega^2\over m_\omega^2} \tr (B^\mu B_\mu)
- \epsilon_2 {g_\omega^2\over 2 m_\omega^2} \tr (B^\mu) \tr ( B_\mu) \cr
& + \bigg[{F_\pi^2 \over 32}(m_\pi^2 + m_\eta^2) \tr (U + U^\da - 2)
+ {\sqrt{3}F_\pi^2 \over 24}(m_\pi^2 - m_K^2) \tr (\lambda_8(U + U^\da))
\bigg],}
\eqno(1)$$
apart from an anomalous contribution to the $\eta'$ mass which is not
relevant here.  We refer to the terms of the lagrangian
by their subscripted forms as given in the first line of this
equation.  Our notation uses the conventional
$L_\mu \equiv U^\da\partial_\mu U$ and the somewhat less usual
definition
$$B^\mu \equiv -{\epsilon^{\mu\alpha\beta\gamma} \over 24\pi^2}
L_\alpha L_\beta L_\gamma.\eqno(2)$$
In most work on skyrmions a trace is immediately taken over this
quantity so that it refers directly to the baryon density, but we
need to retain the distinction between terms of the forms of
${\cal L}_{6,1}$ and ${\cal L}_{6,2}$ (i.e., between terms
involving one or two traces over the six $L_\mu$-operators) which is
at the heart of our source for spin content in the skyrmion.  Further
in eq. (1), $U(\r,t)$ is the U(3) chiral field, $F_\pi$ is the
pion decay constant (with experimental value 186 MeV), and $e$ is
the parameter of the four-derivative term introduced by Skyrme in
order to stabilize the $U$-field.

The first two terms in eq. (1), ${\cal L}_2 + {\cal L}_4,$ are the
original Skyrme lagrangian [2].
The next terms ${\cal L}_{6,1}$ and ${\cal L}_{6,2}$ involve six
derivatives and have been used in the past as possible
$\omega$-coupling repulsive terms
for stabilizing the skyrmion [11,12].  They are equivalent [12] in
SU(2) but not in SU(3), reminiscent of well-known terms with four
derivatives that are equivalent in SU(2) but different in SU(3) [15].
The coefficients of these terms, $\epsilon_1 g_\omega^2/m_\omega^2$ and
$\epsilon_2 g_\omega^2/2 m_\omega^2,$ are taken in a form that allows
easy comparison with the $\omega NN$ coupling constant:
we keep $\epsilon_1 + \epsilon_2 = 1,$ absorbing overall strength
into $g_\omega^2/m_\omega^2,$ with $m_\omega = 782$ MeV, so that
${\cal L}_{6,1} + {\cal L}_{6,2}$ at the SU(2) level continues to give
the usual $\omega$ coupling.  The flavor symmetry breaking term
${\cal L}_{SB}$ in eq. (1) is well known [13] to be important for work
with the SU(3) skyrmion.
Last, we note that in eq. (1) we have not
retained two further terms involving four derivatives of $U$
that were entertained in ref. [10] but then dropped there when they
proved to have only small impact on the issue of spin content;
this omission also avoids problems of terms of fourth order in field
time derivatives which otherwise arise.

The proton spin content is generated from ${\cal L}$ of eq. (1)
by introducing the U(3) matrix
$$U = \exp\bigg[{2 i \over F_\pi}\bigg(\eta'+ \sum_{a=1,8} \lambda_a
\phi_a\bigg)\bigg],\eqno(3)$$
where $\phi_a$ is the pseudoscalar octet and $\eta'$ is the ninth
pseudoscalar meson.  It is well known [5,6] that there is no
contribution to the spin content from a U(1) axial current that is a
complete four-derivative since this vanishes in producing $\Delta\Sigma$
as an integral over all space.  Here we construct the axial current out
of a term in the lagrangian of the form [6]
$${\cal L}' = (2/F_\pi) \partial_\mu\eta' J^\mu\eqno(4)$$
generated by varying $\eta'.$  The static hedgehog makes no
contribution to spin content [6] and thus to evaluate $J^\mu$ we use
collective coordinates for the time dependence [16]
$$U(\vec r,t) = A(t) U_0(\vec r) A^\da(t),\eqno(5)$$
with the hedgehog embedded in SU(3) as
$$U_0 = \exp[i \vec\lambda\cdot\hat{\vec r} F(r)],\eqno(6)$$
where $F(r)$ is the profile function.  The spin content is then
$$\Delta\Sigma = 2 \la\pup|J^3|\pup\ra,\eqno(7)$$
and the contribution of ${\cal L}_{6,1}$ to this is [10]
$$\la\pup|J_{6,1}^3|\pup\ra = - {1 \over 2(3\pi)^3}
 {g_\omega^2 \over m_\omega^2} {\epsilon_1 \over \beta^4}I,\eqno(8)$$
where
$$ I \equiv \int_0^\infty r dr \sin^2 {F \over 2} \sin F
\bigg(F'^2 - F' {\sin F \over r}
+ {\sin^2 F \over r^2}\bigg).\eqno(9)$$
The quantity $1/\beta^2$ is the moment
of inertia for the SU$_{\rm L}$(3) Casimir operator $C_2$(SU$_{\rm L}$(3)).
It appears also in the mass expression [13]
$$M = M_{\rm cl} +
\half\bigg({1 \over \alpha^2} - {1 \over \beta^2}\bigg)
C_2({\rm SU}_{\rm R}(2)) - {3 \over 8 \beta^2} +
 {1 \over 2 \beta^2}C_2({\rm SU}_{\rm L}(3))
+\half \gamma (1 - D^{(8)}(A)),\eqno(10)$$
along with $\alpha^2,$ the moment of inertia for
the SU$_{\rm R}$(2) Casimir operator $C_2$(SU$_{\rm R}$(2)).
Expressions for both moments of inertia are shown in ref. [13].  To
the forms shown there we have now added the additional pieces yielded
[10] by ${\cal L}_{6,1} + {\cal L}_{6,2}.$
The other terms in eq. (10) involve the skyrmion mass $M_{\rm cl}$ and the
SU(3) symmetry breaking term with coefficient $\gamma.$  Expressions for
these quantities are also given in
ref. [13].  Last, forms for the baryon charge radii,
axial couplings, and magnetic moments are provided by Kanazawa [17] for the
minimal Skyrme lagrangian ${\cal L}_2 + {\cal L}_4.$  The introduction
of the terms ${\cal L}_{6,1} + {\cal L}_{6,2}$ in eq. (1) leads to
additional contributions to these observables which are shown in the
appendix to the present paper.  We note that in the present work the
calculation of the observables includes the effects of symmetry breaking
[13].

Before presenting numerical results it is worth noting that the integral
$I$ appearing in the spin content, eqs. (8) and (9), is an approximate
topological constant in the sense that it can be written as
$$\eqalign{I & = \int_0^\infty r dr \sin^2 {F \over 2} \sin F
\bigg[\bigg(F' + {\sin F \over r}\bigg)^2 - 3 F' {\sin F \over r}\bigg] \cr
& = {3\pi \over 4} + \int_0^\infty r dr \sin^2 {F \over 2} \sin F
\bigg(F' + {\sin F \over r}\bigg)^2\cr
& \ge 3\pi/4,}\eqno(11)$$
where we have used the profile function boundary values $F(0) = \pi$ and
$F(\infty) = 0$ and also exploited the fact that $0 < F(r) \le \pi$ for
actual one-baryon profiles.  Since the profile function is near these
values for a good part of the range of integration one might expect from
Taylor-series expansion that the combination $F' + \sin F/r$
in eq. (11) will be small on average, and indeed
carrying out the integration numerically for a wide variety of functions
that satisfy the boundary conditions of $F(r)$ and, like it, fall
monotonically from $r$ near the origin to infinity suggests that the
correction to
$I = 3\pi/4$ is generally less than 10\%.  The case $I = 3\pi/4$ is achieved
if and only if $F' = -\sin F/r$ whose solution $F(r) = 2 \arctan (c/r),$
where $c$ is a constant, has the form suggested by an analysis [18]
of the skyrmion in terms of adiabatic invariants for large baryon number.
As this form is found [18] to
be a reasonable approximation even for $B = 1,$ it is not surprising that
numerically $I \approx 3\pi/4$ for all the cases considered below.  This
in turn allows us to approximate the overall spin content by
$$\Delta\Sigma \approx - {1 \over (3\pi)^3} {g_\omega^2 \over m_\omega^2}
{\epsilon_1 \over \beta^4} {3\pi \over 4}.\eqno(12)$$
Since the moment of inertia $1/\beta^2$ as derived from the lagrangian of
eq. (1) is moderately sensitive to the value for $\epsilon_1$ one should not
conclude from this that $\epsilon_1$ is merely a multiplicative factor in
the spin content.  As $1/\beta^2$ is fixed
by the baryon mass spectrum our fitting procedure largely reduces to
extracting  $1/\beta^2$ and subsequently $\epsilon_1$ from the spectrum
and then determining the spin content from eqs. (7)--(9).
\v
\c{\bf 3. Results and discussion}

In fitting the parameters of eq. (1)
to data we keep $\epsilon_1 + \epsilon_2 = 1$ and take
$g_\omega^2/4\pi = 10,$ as implied by [12] $\omega NN$ coupling and by [11]
the decay $\omega \rightarrow 3\pi.$  The value $m_K = 640$ MeV generally
used in the
symmetry-breaking term of eq. (1) is, as usual [13], larger than the
experimental one, most likely because of the omission of other possible
symmetry-breaking terms [15].
We now vary $F_\pi,$ $e,$ and $\epsilon_1$ so as to produce reasonable
values for baryon masses.  In the previous study [10] it was found that
$F_\pi = 130$ MeV and the rather high value $e = 20$ led to reasonable
$N$ and $\Delta$ masses
and SU(2) nucleon electroweak properties; we then arbitrarily chose
$\epsilon_1 = -0.7,$ which yields\footnote{$^1$}{A somewhat
different
value was shown in ref. [10] due to a numerical error there.}
 $\Delta\Sigma = 0.43$ but very high
masses for the strange baryons.  The results of the current fitting
procedure for the baryon octet and decuplet masses---after subtraction of
the zero-point mass according to the procedure of Yabu and Ando
[13]---are shown in table 1 along with the experimental values.
In table 2 we show the results for the magnetic moments of the baryon
octet.  These fits are generally quite good, certainly of the quality
usually found with skyrmions.  All but the last set of parameters suffer
from a value for the pion decay constant $F_\pi$
which is less than 50 percent of the experimental $F_\pi = 186$ MeV.
This tendency of the skyrmion to require
quite low values for $F_\pi$ is well known [13,16,17].  Case 6 in the
tables refers
to a fit that holds  $F_\pi$ at its physical value (and uses $m_K = 540$
MeV); the necessary subtraction for the mass spectrum in this case is
correspondingly larger [17], and even then the spectrum is rather less
satisfactory than for the cases with $F_\pi$ smaller than 186 MeV.
This case has the further drawback that it gives $g_A = 1.72$ for
the proton, which is considerably worse than the values $g_A \sim 1.24 \pm
0.05$ obtained for cases 1 through 5.  All the cases considered here fall
within the rough range for the parameter $e$ provided [20] by $\pi\pi$
scattering, namely, $e \approx 5 \pm 2.$
In tables 1 and 2 we show numbers for the root of the summed squared
deviations of the
calculated from the measured values for the masses and the magnetic
moments, respectively.  From these values it emerges that, with
respect both to masses and to magnetic moments, the best fits
are obtained for cases 1 and 4 in the tables.  These two fits are not
appreciably different
from each other, and are both substantially better than the other fits shown.

Table 3 gives values for spin content for the six parameter sets we have
considered, as well as providing the moments of inertia $1/\alpha^2$ and
$1/\beta^2.$  The two preferred fits of tables 1 and 2, namely, cases 1
and 4, have similar values for $1/\beta^2$ of around $555 \pm 20$ MeV,
and yield $\Delta\Sigma = 0.17$ and 0.14, respectively.  Inspection of
table 3 leads us to conclude that, while we achieve only
loose bounds on $\epsilon_1,$ these suffice to localize
$\Delta\Sigma$ as lying, in this model, between 0.0 and 0.6, with a
preference for $\Delta\Sigma \sim 0.15.$
\v\v
This research was supported in part by the Israel Science Foundation
and in part by the Yuval Ne'eman Chair in Theoretical Nuclear Physics
at Tel Aviv University.  We are grateful to John Ellis, Marek Karliner,
Jechiel Lichtenstadt, and Andreas Sch\"afer for discussions pertaining to it.
\ej
\c{\bf Appendix}

We quote here the additional terms for baryon observables that must be
added to the expressions of Kanazawa [17] in order to include the effects
of ${\cal L}_{6,1} + {\cal L}_{6,2}$; the notation here follows that of ref.
[17].  To $\Lambda(x)$ and $G(x)$ of eq. (21) there must be added
$$\Delta\Lambda(x) = {1 \over \pi^4}{g_\omega^2 \over m_\omega^2} F_\pi^2
e^4(\epsilon_1 + \epsilon_2)S^4 \dot F^2,\eqno(\A .1)$$
and
$$\Delta G(x) = {1 \over 9 \pi^4}{g_\omega^2 \over m_\omega^2} F_\pi^2
e^4\epsilon_1 x^2 (1 - C) {S^2 \over x^2}
\bigg(2\dot F^2 + {S^2 \over x^2}\bigg).\eqno(\A .2)$$
Last, $E$ of eq. (24) must be modified by
$$\Delta E = {1 \over \pi^4}{g_\omega^2 \over m_\omega^2} F_\pi^2
e^4(\epsilon_1 + \epsilon_2) \int dx x^2 \bigg[\dot F {S^2 \over x^2}
\bigg({S^2 \over x^2} + 2 S C {\dot F \over x}\bigg)\bigg].\eqno(\A .3)$$

\v\v\v\v

\c{\bf References}
\v
\baselineskip 12pt
\parskip 0pc
\parindent 1pc
\hangindent 2pc
\hangafter 10

\item{[1]}  EMC, J. Ashman et al., Phys. Lett. B 206 (1988) 364; Nucl. Phys.
B 328 (1989) 1.
\v
\item{[2]}  T.H.R. Skyrme, Proc. Roy. Soc. London, Series A, 260 (1961)
127; 262 (1961) 237 and Nucl. Phys. 31 (1962) 556.
\v
\item{[3]}  S.J. Brodsky, J. Ellis, and M. Karliner, Phys. Lett. B 206
(1988) 309.
\v
\item{[4]} SMC,  D. Adams et al., Phys. Lett. B 329 (1994) 399.
\v
\item{[5]}  Z. Ryzak, Phys. Lett. B 217 (1989) 325; B 224 (1989) 450.
\v
\item{[6]}  T.D. Cohen and M.K. Banerjee, Phys. Lett. B 230 (1989) 129.
\v
\item{[7]}  R. Johnson, N.W. Park, J. Schechter, V. Soni, and H. Weigel,
Phys. Rev. D 42 (1990) 2998.
\v
\item{[8]}  J. Schechter, A. Subbaraman, and H. Weigel,
Phys. Rev. D 48 (1993) 339.
\v
\item{[9]}  J. Schechter, V. Soni, A. Subbaraman, and H. Weigel,
Mod. Phys. Lett. A 7 (1992) 1.
\v
\item{[10]} G. K\"albermann, J.M. Eisenberg, and A. Sch\"afer, Phys. Lett.
B 339 (1994) 211.
\v
\item{[11]}  G.S. Adkins and C.R. Nappi, Phys. Lett. B 137 (1984) 251.
\v
\item{[12]}  A. Jackson, A.D. Jackson, A.S. Goldhaber, G.E. Brown, and
L.C. Castillejo, Phys. Lett. 154B (1985) 101.
\v
\item{[13]}  H. Yabu and K. Ando, Nucl. Phys. B 301 (1988) 601.
\v
\item{[14]}  H.W. Fearing and S. Scherer, unpublished, hep-ph/9408346.
\v
\item{[15]} G. Pari, B. Schwesinger, and H. Walliser, Phys. Lett. B 255
(1991) 1.
\v
\item{[16]} G.S. Adkins, C.R. Nappi, and E. Witten, Nucl. Phys. B 228
(1983) 552.
\v
\item{[17]} A. Kanazawa, Prog. Theor. Phys., 77 (1987) 1240.
\v
\item{[18]} G. K\"albermann and J.M. Eisenberg, J. Phys. G  13 (1987)
1029.
\v
\item{[19]}  Particle Data Group, L. Montanet et al., Phys. Rev. D 50 (1994)
1173.
\v
\item{[20]} J.F. Donoghue, E. Golowich, and B.R. Holstein, Phys. Rev.
Lett. 53 (1984) 747.
\ej
\c{TABLE 1}

\c{Parameters for ${\cal L}$ of eq. (1) and corresponding masses}
\v
\hrule
\settabs\+&\hskip 1.0 true cm &\hskip 1.2 true cm &\hskip 1.8 true cm
&\hskip 1.2 true cm &\hskip 2.0 true pc&\hskip 1.7 true cm &\hskip 1.7 true cm
&\hskip 1.7 true cm &\hskip 1.7 true cm &\hskip 1.0 true pc &
&\hskip 1.7 true cm &\cr
\v
\+&Case&&Parameters&&&Masses [MeV] (octet/decuplet)&&&&&
$\sqrt{\sum(M_{\rm th}-M_{\rm exp})^2}$&\cr
\+&&\hrf&\hrf&\hrf&&\hrf&\hrf&\hrf&\hrf&&&\cr
\+&&$e$&${F_\pi \over F_\pi({\rm exp})}$&$\epsilon_1$
&&$N_8$/$\Delta_{10}$&$\Lambda_8$/$\Sigma_{10}$&$\Sigma_8$/$\Xi_{10}$
&$\Xi_8$/$\Omega_{10}$&&&\cr
\v
\hrule
\v
\+&1&6.0&0.44&-1.00&&938&1111&1240&1366&&98&\cr
\+&&&&&&1226&1336&1485&1649&&&\cr
\vskip 0.3 true pc
\+&2&6.0&0.40&-1.50&&936&1114&1246&1375&&120&\cr
\+&&&&&&1234&1338&1469&1624&&\cr
\vskip 0.3 true pc
\+&3&6.0&0.48&-0.35&&928&1124&1271&1402&&131&\cr
\+&&&&&&1233&1342&1488&1659&&\cr
\vskip 0.3 true pc
\+&4&6.6&0.44&-0.71&&933&1108&1238&1364&&99&\cr
\+&&&&&&1249&1351&1480&1633&&\cr
\vskip 0.3 true pc
\+&5&5.4&0.44&-1.26&&919&1119&1272&1410&&159&\cr
\+&&&&&&1203&1317&1469&1643&&\cr
\vskip 0.3 true pc
\+&6&8.0&1.00&-1.45&&939&1120&1253&1387&&120&\cr
\+&&&&&&1236&1344&1476&1634&&\cr
\vskip 0.3 true pc
\+&exp$^*$&&&&&939&1116&1193&1318&&&\cr
\+&&&&&&1232&1385&1530&1672&&&\cr
\vskip 0.5 pc
\hrule
\vskip 0.5 pc
\noindent $^*$Ref. [19]
\ej
\c{TABLE 2}

\c{Parameters for ${\cal L}$ and octet magnetic moments}
\v
\hrule
\settabs\+&\hskip 1.2 true cm &\hskip 1.5 true cm
&\hskip 1.5 true cm &\hskip 1.5 true cm
&\hskip 1.5 true cm &\hskip 1.5 true cm &\hskip 1.5 true cm
&\hskip 1.5 true cm &\hskip 1.5 true cm
&\hskip 0.0 true pc &\hskip 1.7 true cm &\cr
\v
\+&Case&Magnetic moments [$\mu_N$]&&&&&
&&&$\sqrt{\sum(\mu_{\rm th}-\mu_{\rm exp})^2}$&\cr
\+&&\hrf&\hrf&\hrf&\hrf&\hrf&\hrf&\hrf&\hrf&&&\cr
\+&&p&n&$\Lambda$&$\Sigma^+$&$\Sigma^0$&$\Sigma^-$&$\Xi^0$&$\Xi^-$&&\cr
\v
\hrule
\v
\+&1&2.65&-1.73&-0.83&2.80&0.84&-1.10&-1.75&-0.89&&0.73&\cr
\vskip 0.3 true pc
\+&2&2.80&-1.77&-0.90&2.98&0.93&-1.12&-1.86&-0.99&&0.93&\cr
\vskip 0.3 true pc
\+&3&2.54&-1.72&-0.75&2.64&0.75&-1.13&-1.65&-0.79&&0.58&\cr
\vskip 0.3 true pc
\+&4&2.49&-1.57&-0.77&2.65&0.81&-1.02&-1.62&-0.88&&0.69&\cr
\vskip 0.3 true pc
\+&5&2.87&-1.98&-0.88&3.01&0.88&-1.24&-1.90&-0.89&&0.93&\cr
\vskip 0.3 true pc
\+&6&2.33&-1.76&-0.78&2.42&0.72&-0.97&-1.65&-0.65&&0.68&\cr
\vskip 0.3 true pc
\+&exp$^*$&2.79&-1.91&-0.61&2.46&&-1.16&-1.25&-0.65&\cr
\vskip 0.5 pc
\hrule
\vskip 0.5 pc
\noindent $^*$Ref. [19]
\v\v\v\v
\c{TABLE 3}

\c{The moments of inertia $1/\alpha^2$ and $1/\beta^2$, $I$ of eq. (9),
and the spin content $\Delta\Sigma$}
\v
\hrule
\settabs\+&\hskip 1.5 true cm &\hskip 3.0 true cm &\hskip 3.0 true cm
&\hskip 2.5 true cm &\hskip 2.5 true cm &\cr
\v
\+&Case&$1/\alpha^2$ [MeV]&$1/\beta^2$ [MeV]&$I$&$\Delta\Sigma$&\cr
\v
\hrule
\v
\+&1&130&536&2.47&0.17&\cr
\vskip 0.3 true pc
\+&2&132&615&2.47&0.34&\cr
\vskip 0.3 true pc
\+&3&130&452&2.47&0.04&\cr
\vskip 0.3 true pc
\+&4&145&574&2.47&0.14&\cr
\vskip 0.3 true pc
\+&5&115&473&2.46&0.17&\cr
\vskip 0.3 true pc
\+&6&130&810&2.49&0.58&\cr
\vskip 0.3 true pc
\+&exp$^*$&&&&$0.27\pm 0.13$&\cr
\vskip 0.5 pc
\hrule
\vskip 0.5 pc
\noindent $^*$Ref. [4]

\bye